\documentclass[12pt]{iopart}
\usepackage{iopams}
\usepackage{graphicx}
\begin{document}
%

\title{Do we live in a ``small Universe''?}

\author{Ralf Aurich$^1$, Holger S.\ Janzer$^1$, Sven Lustig$^1$, and Frank Steiner$^1$}

\address{$^1$Institut f\"ur Theoretische Physik, Universit\"at Ulm,\\
Albert-Einstein-Allee 11, D-89069 Ulm, Germany}

\begin{abstract}
We compute the effects of a compact flat universe on the
angular correlation function, the angular power spectrum,
the circles-in-the-sky signature,
and the covariance matrix of the spherical harmonics coefficients of the
cosmic microwave background radiation using the full Boltzmann physics.
Our analysis shows that the Wilkinson Microwave Anisotropy Probe (WMAP)
three-year data are well compatible with the possibility
that we live in a flat 3-torus with volume
$\simeq 5 \cdot 10^3 \hbox{Gpc}^3$.
\end{abstract}



\section{Introduction}

At present, all data are consistent with, and in fact strongly  
support, the standard big bang model based on general relativity and  
the cosmological principle leading to the general class of  
Friedmann-Lema\^itre universes possessing the space-time structure  
$R\times M^3$.
Here $R$ describes cosmic time and $M^3$ the  
three-dimensional comoving spatial section of constant curvature
$K = 0$ or $K = \pm1$.
Since the Einstein gravitational field equations are  
differential equations, they govern the local properties of space-time  
but not the global geometry of the Universe at large, i.\,e.\ they do  
not determine the spatial curvature and topology, and thus the shape  
of the Universe.

Although a full quantum theory of gravity has not yet been  
established, it is commonly assumed that our Universe emerged from  
quantum fluctuations during the Planck era and that its spatial  
curvature and topology are invariants,
i.\,e.\ have not changed ever since.
The spatial curvature $K$ can be inferred from astronomical  
observations of the total energy density $\epsilon_{\hbox{\scriptsize tot}}$
of the Universe  (at the present epoch) which is usually expressed in terms  
of the dimensionless density parameter
$\Omega_{\hbox{\scriptsize tot}} = \epsilon_{\hbox{\scriptsize tot}}/
\epsilon_{\hbox{\scriptsize crit}}$.
($K = 0$ for $\Omega_{\hbox{\scriptsize tot}}= 1,$ $K = \pm 1$ for
$\Omega_{\hbox{\scriptsize tot}} \gtrless 1$.
$\epsilon_{\hbox{\scriptsize crit}} = 3H^{2}_{0}c^2/(8\pi G)$,  
where $H_0 = 100 h (\hbox{km}/\hbox{s})/\hbox{Mpc}$
denotes the Hubble constant.)

Measurements of the temperature anisotropy of the cosmic microwave  
background (CMB) by NASA's satellite Wilkinson Microwave Anisotropy  
Probe (WMAP) suggest (using the 3-year data combined with other  
observations) $\Omega_{\hbox{\scriptsize tot}} \simeq 1.0$
\cite{Spergel_et_al_2007}.
Since this value for the density parameter is compatible with
Euclidean geometry, we shall assume in this paper that our Universe is,
indeed, flat.
(For a discussion of positively or negatively curved universes, see e.\,g.\  
\cite{Luminet_Weeks_Riazuelo_Lehoucq_Uzan_2003,Aurich_Lustig_Steiner_2004c,%
Aurich_Lustig_Steiner_2005a} and
\cite{Aurich_Steiner_2000,Aurich_Lustig_Steiner_Then_2004a,%
Aurich_Lustig_Steiner_Then_2004b}, respectively.)
Having thus fixed the curvature, there  
remains the question of the topology of our Universe.

It is a mathematical fact that fixing the curvature $K$ does not  
determine uniquely the topology and thus the shape of the universe.  
Only if it is {\it assumed} that the spatial section is simply  
connected, it follows that the topology is (in the flat case) given by  
the infinite Euclidean 3-space $E^3$.
Exactly this assumption is made in the popular $\Lambda$CDM model
\cite{Spergel_et_al_2007} which is composed of radiation (r),
ordinary baryonic matter (b), cold dark matter (cdm) and has
a cosmological constant $\Lambda$, i.\,e.\,
$\Omega_{\hbox{\scriptsize tot}} =
\Omega_{\hbox{\scriptsize r}} + \Omega_{\hbox{\scriptsize b}} +
\Omega_{\hbox{\scriptsize cdm}} + \Omega_\Lambda \equiv 1$.

It is the purpose of this paper to demonstrate that the WMAP 3yr data
\cite{Spergel_et_al_2007} are well compatible with the possibility
that we live in a ``small Universe'' having the shape of a flat 3-torus
whose fundamental domain is a cube with side length $L\simeq 4$
corresponding to a volume of $\simeq 5 \cdot 10^3 \hbox{Gpc}^3$.
($L$ is given in units of the Hubble length
$L_{\hbox{H}} = c/H_0 \simeq 4.26 \hbox{Gpc}$ for $h=0.704$.)
For previous works on a toroidal universe, see \cite{%
Schwarzschild_1900, %
Zeldovich_1973, %
Fang_Sato_1983, %
Ellis_Schreiber_1986, %
Fang_Mo_1987, %
Sokolov_1993, %
Starobinsky_1993, %
Stevens_Scott_Silk_1993, %
Jing_Fang_1994, %
deOliveira-Costa_Smoot_1995, %
deOliveira-Costa_Smoot_Starobinsky_1996, %
Levin_Scannapieco_Silk_1998, %
Scannapieco_Levin_Silk_1999, %
Inoue_2001, %
Cornish_Spergel_Starkman_Komatsu_2003, %
Riazuelo_Uzan_Lehoucq_Weeks_2004, %
Riazuelo_et_al_2004, %
Phillips_Kogut_2006, %
Kunz_et_al_2006, %
Kunz_et_al_2007}.
It turns out that the torus model describes the data much better than the
best-fit $\Lambda$CDM model since it exhibits the suppression of the
CMB anisotropy at large scales first observed by COBE
\cite{Hinshaw_et_al_1996}.

Let us point out that Linde \cite{Linde_2004}
has argued recently that a compact flat universe like the toroidal universe
provides the simplest way to solve the problem of initial conditions
for the low-scale inflation.
While the quantum creation of a closed or an infinite open inflationary
universe is exponentially suppressed, there is no such suppression
for the toroidal universe.

In \cite{Spergel_et_al_2007} the WMAP team offers several sets of
cosmological parameters depending on which cosmological data sets are taken
into account.
In this paper we use the cosmological parameters of the $\Lambda$CDM model
of Table 2 in \cite{Spergel_et_al_2007} based on all astronomical
observations, see their column ``3 Year + ALL Mean'', i.\,e.\
$\Omega_{\hbox{\scriptsize b}} = 0.044$,
$\Omega_{\hbox{\scriptsize cdm}} = 0.223$,
$\Omega_\Lambda = 0.733$, $h=0.704$, $n_s = 0.947$, $\tau=0.073$.
The distance to the surface of last scattering (SLS) is
$L_{\hbox{\scriptsize SLS}} = \Delta \eta L_{\hbox{H}} \simeq 14.2 \hbox{Gpc}$
where $\Delta \eta =\eta_0-\eta_{\hbox{\scriptsize SLS}} = 3.329$
($\eta$ is the conformal time).

The temperature fluctuations $\delta T(\hat n)$ of the CMB are calculated
using the full Boltzmann physics of the coupled baryon-photon fluid,
i.\,e.\ the ordinary and the integrated Sachs-Wolfe effect,
the Doppler contribution, the Silk damping and
the reionization are taken into consideration.
The reionization is taken into account by modifying the
ionization fraction $x_e$ computed with recfast
\cite{Seager_Sasselov_Scott_1999}
\begin{equation}
x_e = x_e^{\hbox{\scriptsize recfast}} +
\left\{1+\frac{Y_p}{2Y_p(1-Y_p)}\right\}
\frac{\hbox{erfc}(\alpha(z-\beta))}2
\end{equation}
by adding a term proportional to an erfc-function with parameters $\alpha$ and $\beta$.
(Here $Y_p=0.24$ denotes the Helium abundance.)
We set $\alpha = 0.4$ and $\beta=9$ which leads to an optical depth
of $\tau=0.073$.

The crucial difference between the toroidal and the simply connected
$\Lambda$CDM model arises from the discrete spectrum of the vibrational modes
$\{ \vec k =
\frac{2\pi}L \vec \nu; \vec \nu =(\nu_x,\nu_y,\nu_z) \in {\mathbb Z}^3\}$
due to the periodic boundary conditions imposed on the regular solutions
of the Helmholtz wave equation for the 3-torus.
(Note that a given wave number $k=|\vec k\,|$ contributes with a
highly irregular multiplicity determined by the number of representations
of $a\in {\mathbb N}$ by a sum of 3 squares of integers.)
Since the 3-torus does not support wave lengths longer than $L$,
there is an infrared cutoff $k\geq 2\pi/L$
(the mode $k=0$ is not included being a pure gauge mode)
which leads to the observed suppression of CMB power at large scales
as we shall show now.
In our calculations we use the exact mode spectrum for the
first 10\,000 eigenvalues, i.\,e.\ the first 5\,503\,602 eigenmodes.
The contribution of the higher modes is taken into account
by a smooth remainder term as in the case of an infinite universe.

\section{Large scale temperature correlations and the angular power spectrum}

\begin{figure}[htb]
\begin{center}
\hspace*{-10pt}\begin{minipage}{9cm}
\vspace*{-25pt}\includegraphics[width=9.0cm]{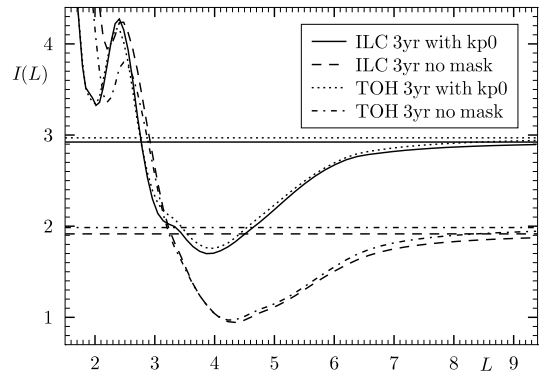}
\end{minipage}
\vspace*{-30pt}
\end{center}
\caption{\label{Fig:C_theta_Integ_ns_0947}
The integrated weighted temperature correlation difference $I(L)$
is shown for the torus model in dependence on the length $L$
and for the best-fit $\Lambda$CDM model (straight horizontal lines)
as described in \cite{Spergel_et_al_2007}
incorporating the cosmological parameters of the ``WMAP3yr + ALL'' model.
The models are compared with four data sets derived from the WMAP 3yr data.
}
\end{figure}

Let us consider the two-point temperature correlation function
$C(\vartheta) = \left< \delta T(\hat n)\, \delta T(\hat n') \right>$,
$\hat n\cdot  \hat n' = \cos\vartheta$,
where $\left< \dots \right>$ denotes an ensemble average over the
primordial fluctuations, respectively,
an ensemble average over the universal observers.
In Fig.\ \ref{Fig:C_theta_Integ_ns_0947} we show as a function of the
torus length $L$ the integrated weighted temperature correlation difference 
\begin{equation}
I := \int_{-1}^1 d(\cos\vartheta)
\frac{(C^{\hbox{\scriptsize model}}(\vartheta)-
C^{\hbox{\scriptsize obs}}(\vartheta))^2}
{\hbox{Var}(C^{\hbox{\scriptsize model}}(\vartheta))}
\end{equation}
between $C^{\hbox{\scriptsize model}}(\vartheta)$ for the torus model and
$C^{\hbox{\scriptsize obs}}(\vartheta)$ for four data sets derived from
the WMAP 3yr data.
The full and the dotted curves show $I(L)$ derived from the ILC 3yr map
and the TOH 3yr cleaned map \cite{Tegmark_deOliveira_Costa_Hamilton_2003}
both with Kp0 mask, respectively,
while the dashed and dotted-dashed curves refer to the ILC 3yr map
and the TOH 3yr cleaned map both without mask.
The horizontal lines show the corresponding results for the best-fit
infinite $\Lambda$CDM model.
It is seen from Fig.\ \ref{Fig:C_theta_Integ_ns_0947} that $I(L)$
possesses a pronounced minimum at $L=3.86$ for the maps with Kp0 mask,
respectively at $L=4.35$ for the maps without mask, which allows us
to fix the best-fit torus model depending on the used data set.
This shows that the WMAP-ILC 3yr map and the TOH 3yr cleaned map
lead to the same best torus model.
The uncertainty results mainly from the data near the galactic plane,
i.\,e.\ by including them using no mask or
by excluding them by applying the Kp0 mask.
However, in both cases, the torus length $L$ is close to $L=4$,
such that the main result is not affected by the foregrounds near the
galactic plane.
The values of $I(L)$ obtained for the best-fit models are substantially
lower than the corresponding values for the best-fit $\Lambda$CDM model
demonstrating the suppressed CMB power of the toroidal universe.
Obviously, if the toroidal universe gets too large $(L\gtrsim 9)$,
the results are very similar to the infinite flat universe.

\begin{figure}[thb]
\begin{center}
\begin{minipage}{9cm}
\vspace*{-25pt}\includegraphics[width=9.0cm]{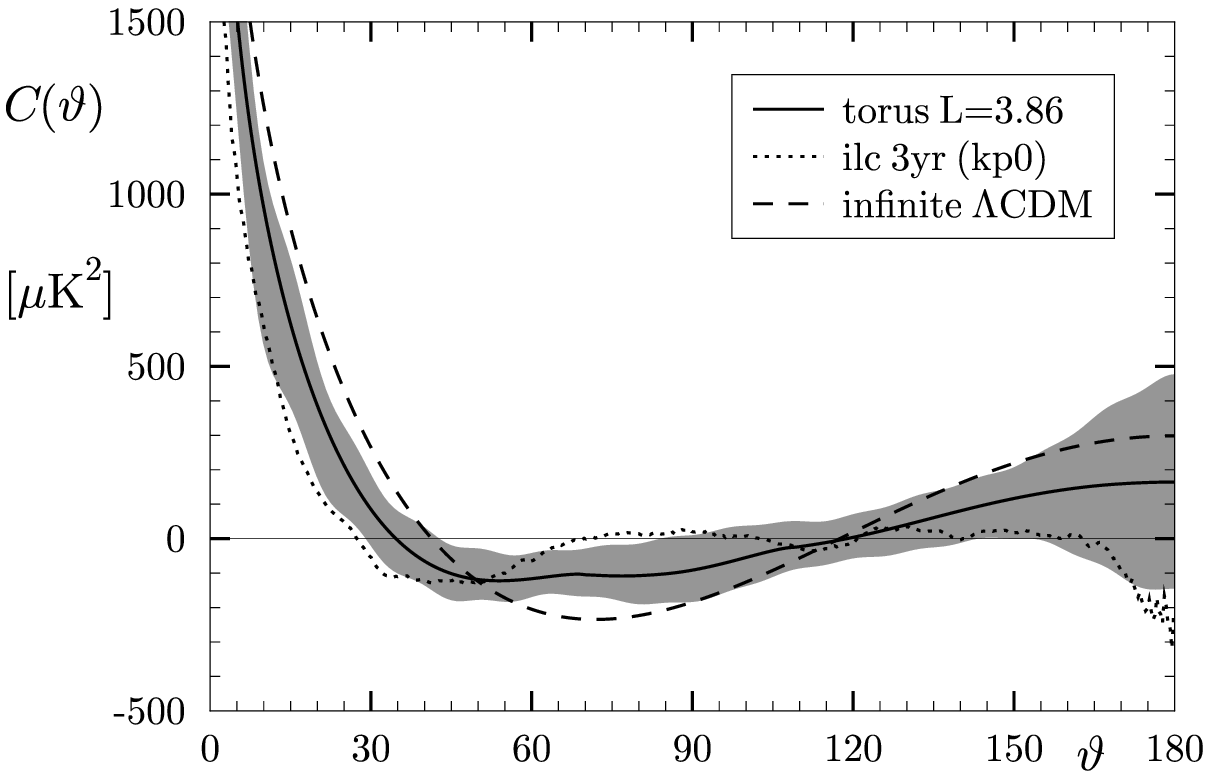}
\end{minipage}
\vspace*{-30pt}
\end{center}
\caption{\label{Fig:C_theta}
The temperature correlation function $C(\vartheta)$
is shown for the torus model with the length $L= 3.86$
and for the best-fit $\Lambda$CDM model.
The grey band shows the $1\sigma$ deviation computed from
10\,000 simulations of the  $L= 3.86$ torus model.
}
\end{figure}

\begin{figure}[htb]
\begin{center}
\begin{minipage}{9cm}
\vspace*{-25pt}\includegraphics[width=9.0cm]{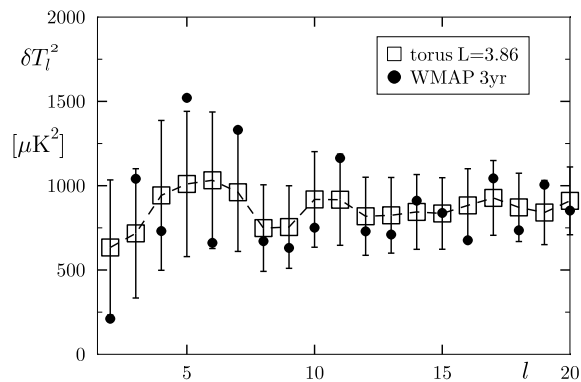}
\end{minipage}
\vspace*{-30pt}
\end{center}
\caption{\label{Fig:C_l}
The angular power spectrum $\delta T_l^2$
is shown for the torus model with the length $L= 3.86$
with the cosmic variance in comparison with the WMAP 3yr data.
}
\end{figure}

In Fig.\ \ref{Fig:C_theta} we show the mean temperature correlation function
$C(\vartheta)$ for the torus model with the length $L= 3.86$ (full line)
together with the $1\sigma$ standard deviation (grey band).
The $1\sigma$ band is obtained from 10\,000 sky map simulations of
the considered torus model.
For each realization the correlation function $C(\vartheta)$ is computed
which in turn gives the standard deviation for this sample.
This is compared to the WMAP-ILC 3yr (Kp0) data (dotted line)
and the best-fit $\Lambda$CDM model (dashed line).
As expected, the torus model is in much better agreement with the
data than the $\Lambda$CDM model.
Notice that the difference
$C^{\hbox{\scriptsize obs}}(\vartheta)-
C^{\hbox{\scriptsize model}}(\vartheta)$
between the data and the torus model is almost
always a factor two smaller than the difference between the data and
the $\Lambda$CDM model.
In Fig.\ \ref{Fig:C_l} we compare the power spectrum
$\delta T_l^2 = l(l+1) C_l / (2\pi)$ (including cosmic variance)
for the same torus model with the WMAP 3yr power spectrum
\cite{Spergel_et_al_2007}, which is produced by combining the
maximum likelihood estimated spectrum from $l = 2-10$ with the
pseudo-$C_l$ based cross-power spectra for $l > 10$.
This WMAP 3yr power spectrum matches the power obtained from
the previous sky maps without using any masks.
Since we are mainly concerned with the large scale structure,
we present only the low multipoles with $2 \leq l \leq 20$.
We have checked, however, that the power spectrum reproduces correctly the 
higher acoustic peak structure.
The overall normalization is the only free parameter and has been determined
by adjusting the power spectrum in the range $20 \leq l \leq 100$
to the WMAP data.

\begin{figure}[htb]
\begin{center}
\begin{minipage}{9cm}
\vspace*{-25pt}\includegraphics[width=9.0cm]{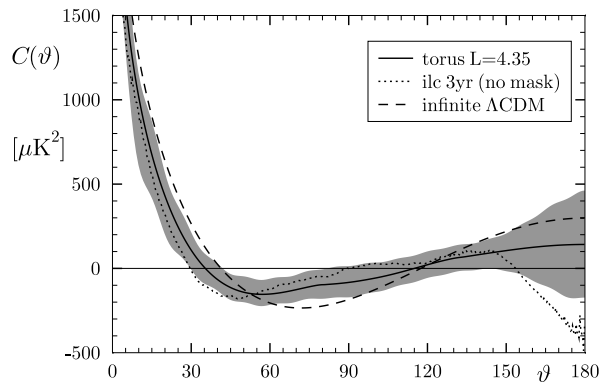}
\end{minipage}
\vspace*{-30pt}
\end{center}
\caption{\label{Fig:C_theta_no_mask}
The temperature correlation function $C(\vartheta)$
is shown for the torus model with the length $L= 4.35$
and for the best-fit $\Lambda$CDM model.
The grey band shows the $1\sigma$ deviation computed from
10\,000 simulations of the  $L= 4.35$ torus model.
}
\end{figure}

\begin{figure}[htb]
\begin{center}
\begin{minipage}{9cm}
\vspace*{-35pt}\includegraphics[width=9.0cm]{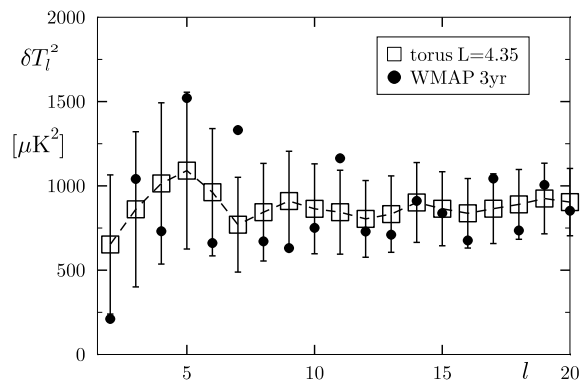}
\end{minipage}
\vspace*{-30pt}
\end{center}
\caption{\label{Fig:C_l_no_mask}
The angular power spectrum $\delta T_l^2$
is shown for the torus model with the length $L= 4.35$
with the cosmic variance in comparison with the WMAP 3yr data.
}
\end{figure}

In Fig.\ \ref{Fig:C_theta_no_mask} and Fig.\ \ref{Fig:C_l_no_mask}
a similar comparison is carried out for the torus model with $L=4.35$
which now refers to the WMAP-ILC 3yr data without mask.
Again a remarkable agreement with the data is found.

\newpage

\section{The circles-in-the-sky signature}

\begin{figure}[htb]
\begin{center}
\begin{minipage}{9cm}
\vspace*{-30pt}\includegraphics[width=9.0cm]{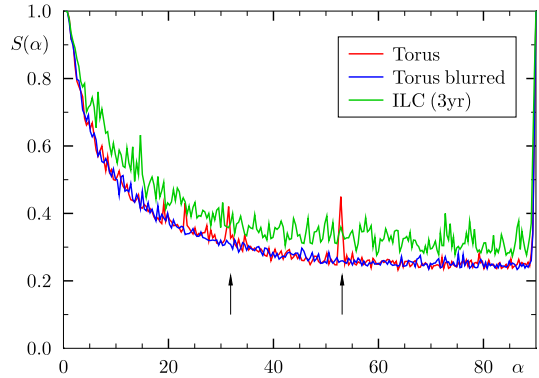}
\end{minipage}
\vspace*{-30pt}
\end{center}
\caption{\label{Fig:S_max_cits}
The circles-in-the-sky signature $S(\alpha)$ is shown in dependence on
the radius $\alpha$ of the back-to-back circles for a sky-map simulation
of a torus with $L=4$ (red) showing peaks at the corresponding
circle radii $\alpha\simeq 31.84^\circ$ and $\alpha\simeq 53.08^\circ$
(indicated by arrows).
Disturbing this sky map by $50\mu\hbox{K}$ destroys this signature (blue).
The result for the ILC-3yr map of the WMAP team is displayed as a green curve
which possesses generally larger values than the simulation.
}
\end{figure}

Up to now we have compared the statistical properties
for the CMB of a torus model with the observations.
There are, however, two quantities,
which can reveal the topology of the Universe more directly.
Besides a signature in the covariance matrix,
to which we devote the next section,
this is the circles-in-the-sky signature
\cite{Cornish_Spergel_Starkman_1998b}.
On each circle pair on the CMB sky which is connected by the
periodicity condition of the topological cell,
there should be similar temperature fluctuations due to the Sachs-Wolfe effect.
Expanding the temperature $\delta T_i(\phi)$ along a circle in a Fourier
series $\delta T_i(\phi)=\sum_m T_{im} e^{\hbox{\scriptsize i}m\phi}$,
$0\leq \phi\leq 2\pi$,
allows one to define the $m$-weighted circle signature for
two circles $i$ and $j$ having a radius $\alpha$
as \cite{Cornish_Spergel_Starkman_Komatsu_2003}
\begin{equation}
\label{Eq:cits}
S_{ij}(\alpha,\beta) \; := \;
\frac{2\sum_m m T_{im} T_{jm}^\star e^{\hbox{\scriptsize i}m\beta}}
{\sum_m m \big(|T_{im}|^2+|T_{jm}|^2\big)}
\hspace{10pt} .
\end{equation}
The angle $\beta$ takes a possible shift between the two circles into account.
In \cite{Cornish_Spergel_Starkman_Komatsu_2003} the WMAP 1yr data are
analysed and no topological signature is found.
In order to investigate the question whether this negative result already
rules out a torus cell as advertised in the previous section,
we produce sky-map simulations for a torus with length $L=4$
using the full Boltzmann physics and taking all eigenmodes up to
the wavenumber $k=384.75$ into account,
i.\,e.\ in total 61,556,892 different wavevectors $\vec k$.
The contributions of the modes are computed up to
$l_{\hbox{\scriptsize max}}=1000$.
A torus model with $L=4$ possesses
3 circle pairs with radius $\alpha\simeq 53.08^\circ$ and
6 circle pairs with radius $\alpha\simeq 31.84^\circ$.
The sky maps are computed in a HEALPix resolution
$N_{\hbox{\scriptsize side}}=512$
and are then smoothed to a resolution of $0.5^\circ$ and
finally downgraded to $N_{\hbox{\scriptsize side}}=128$.
The maximum $S(\alpha) := \max _{ij\beta}S_{ij}(\alpha,\beta)$
of the circle signature (\ref{Eq:cits}) reveals the torus cell clearly
as shown in Fig.\,\ref{Fig:S_max_cits},
where the arrows indicate the peaks corresponding to the two radii.
Now we disturb the sky maps by modifying the spherical expansion coefficients
\begin{equation}
a_{lm} \; \to \;
\frac 1{\sqrt{1+f^2}}\; \Big( a_{lm} \, + \, f \, a_{l,p(m)}\Big) 
\hspace{10pt} ,
\end{equation}
where $f$ is a constant factor and $p(m)$ is a permutation of $m\in[-l,l]$.
With increasing $f$ an  increasingly disturbed sky map is generated
without altering the statistical properties
since only the $l$ subspace is permuted.
We found that the topological peaks in the circle signature $S(\alpha)$ 
vanish if the sky maps are modified by 50$\mu\hbox{K}$ on the average
(see Fig.\,\ref{Fig:S_max_cits}).
Thus the crucial question emerges:
what is the accuracy of the present CMB sky maps,
i.\,e.\ is the CMB accuracy better than 50$\mu\hbox{K}$?
In the latter case the analysis carried out in
\cite{Cornish_Spergel_Starkman_Komatsu_2003}
would rule out the torus model.
The ILC map provided by the WMAP team does not specify the accuracy,
and its usage for cosmological analyses is not recommended.

\begin{table}
  \begin{center}
    \begin{tabular}{|c|c|c|c|}
      \hline
      $N_{\hbox{\scriptsize side}}$& 128 & 256 & 512 \\
      \hline
      &&&\\
      $\langle \sigma(i)\rangle^{\hbox{\scriptsize Kp2}}_{\hbox{\scriptsize 1yr}}$&
$47.3 \pm 15.5$  & $64.2 \pm 20.8$& $73.7 \pm 24.5$ \\
      &&&\\
      \hline
      &&&\\
      $\langle \sigma(i) \rangle^{\hbox{\scriptsize Kp2}}_{\hbox{\scriptsize 3yr}}$
&$26.9 \pm 8.9$  & $36.9 \pm 12.0$& $42.9 \pm 14.5$ \\
      &&&\\
      \hline
    \end{tabular}
\caption{The average scatter (\ref{Eq:Sigma_Pixel}) of the pixels outside
the Kp2 mask is given in $\mu\hbox{K}$ for the 1yr and the 3yr data
with respect to the 8 differencing assemblies.
The standard mean deviation of $\sigma(i)$ from
$\langle \sigma(i)\rangle^{\hbox{\scriptsize Kp2}}$ is also given.
}
    \label{Tab:sigma_wmap}

  \end{center}
\end{table}

Let us at first discuss the detector noise
which is given by the WMAP team \cite{Hinshaw_et_al_2006} for the
sky maps for the 8 differencing assemblies $x$ with
$x \in \{Q_1,Q_2,V_1,V_2,W_1,W_2,W_3,W_4\}$.
We construct from the 8 foreground reduced sky maps $\delta T_{x}$
a mean value per pixel
$\overline{\delta T}(i):=\frac 18 \sum_{x} \delta T_{x}(i)$
where $i$ denotes the pixel number with
$i \in \{0,1,..,12 N_{\hbox{\scriptsize side}}^2-1\}$.
In order to carry out the average, the differencing assemblies $x$ are smoothed
to the common resolution of the $Q_1$ band.
With this mean value one can define a
root mean square deviation per pixel
\begin{equation}
\sigma(i) \; :=\;
\sqrt{\frac{\sum_{x} \Big(\delta T_{x} (i)- \overline{\delta T}(i)\Big)^2}{7}}
\end{equation}
measuring the scatter around the mean value
$\overline{\delta T}(i)$ of a pixel.
From this root mean square deviation $\sigma(i)$
one can define an average over a given domain
which we take as the area outside the Kp2 mask of the 3 year WMAP data
excluding also point sources, i.\,e.\ we define
\begin{equation}
\label{Eq:Sigma_Pixel}
\langle \sigma(i)\rangle^{\hbox{\scriptsize Kp2}} \; := \;
\frac{\sum_s \sigma(s)}{\sum_s 1}
\hspace{10pt} ,
\end{equation}
where the sums run only over the pixels outside the Kp2 mask.
The results are presented in Table \ref{Tab:sigma_wmap}
for three different $N_{\hbox{\scriptsize side}}$ resolutions
for the 1yr and the 3yr sky maps.
In the case of the highest resolution $N_{\hbox{\scriptsize side}}=512$
a mean scatter of $74\mu\hbox{K}$ and $43\mu\hbox{K}$ around the averaged
temperature value is obtained for the 1yr and the 3yr maps, respectively. 
Downgrading the maps to lower values of $N_{\hbox{\scriptsize side}}$
averages neighbouring pixels and thus reduces the detector noise.
The lowest resolution for which a circle signature $S(\alpha)$ can be computed
is  $N_{\hbox{\scriptsize side}}=128$.
Even in this case a mean scatter of $47\mu\hbox{K}$ and $27\mu\hbox{K}$
is obtained.
Since the negative result in \cite{Cornish_Spergel_Starkman_Komatsu_2003}
with respect to a topological detection is based on the one-year data
with a mean scatter of
$\langle \sigma(i)\rangle^{\hbox{\scriptsize Kp2}}_{\hbox{\scriptsize 1yr}}
=47\mu\hbox{K}$ outside the Kp2 mask,
a possible existing signature can with this accuracy easily be masked
as the simulation presented in Fig.\,\ref{Fig:S_max_cits} demonstrates.

\begin{table}
  \begin{center}
    \begin{tabular}{|c|c|c|c|}
     \hline
      frequency & 
 $\Delta_{\hbox{\scriptsize fore}}^{\hbox{\scriptsize 1yr}}$ &
 $\Delta_{\hbox{\scriptsize fore}}^{\hbox{\scriptsize 3yr}}$ &
 $\Delta_{\hbox{\scriptsize 3yr}}^{\hbox{\scriptsize 1yr}}$ \\
     \hline
      $K$ & & & $34.1$  \\
     \hline
      $Ka$ & & & $25.9$  \\ 
     \hline
      $Q$ & $66.3$ & $28.9$ & $21.3$ \\
     \hline
      $V$ & $33.1$ & $6.2$ & $16.7$ \\
     \hline
      $W$ & $50.5$ & $5.3$ & $22.1$ \\
     \hline
    \end{tabular}
\caption{The mean amplitude of the foreground which is taken into account in
the 1yr and 3yr data outside the Kp2 mask is listed in the 2nd and 3rd column.
The mean difference between the 1yr and 3yr data outside the Kp2 mask is given
in the last column.
The values are given in $\mu\hbox{K}$.}
    \label{Tab:DA_foreground_wmap}
  \end{center}
\end{table}

Until now the focus was on the domain outside the Kp2 mask
in order to obtain an estimate for the accuracy of the temperature
pixel values ignoring a problematic foreground contamination.
Within the Kp2 mask which covers 15\% of the sky,
the foreground dominates the CMB signal and an additional uncertainty
with respect to the foreground  removal arises.
In the cleaned 3yr maps ``substantial errors $(\geq 30\mu\hbox{K})$
remain inside the Kp2 cut due to the limitations in the template model''
\cite{Hinshaw_et_al_2006}.
This residual foreground contamination is also the likely reason
why we obtain different favoured lengths $L$ of the torus cell
depending on using the Kp0 mask or not as discussed in the  previous section.
An estimate for the uncertainty in the foreground  removal procedure
is obtained by taking the difference between the primary maps
$\delta T(i)_{\hbox{\scriptsize pri}}$
and the foreground reduced maps  $\delta T(i)_{\hbox{\scriptsize forered}}$
per frequency band provided by the WMAP team,
i.\,e.\ by defining the average difference
$\Delta_{\hbox{\scriptsize fore}}^{\hbox{\scriptsize 1yr}} :=
\langle \delta T(i)_{\hbox{\scriptsize pri}}^{\hbox{\scriptsize 1yr}}-
\delta T(i)_{\hbox{\scriptsize forered}}^{\hbox{\scriptsize 1yr}}
\rangle^{\hbox{\scriptsize Kp2}}$ and
$\Delta_{\hbox{\scriptsize fore}}^{\hbox{\scriptsize 3yr}} :=
\langle \delta T(i)_{\hbox{\scriptsize pri}}^{\hbox{\scriptsize 3yr}}-
\delta T(i)_{\hbox{\scriptsize forered}}^{\hbox{\scriptsize 3yr}}
\rangle^{\hbox{\scriptsize Kp2}}$ outside the Kp2 mask.
One obtains from Table \ref{Tab:DA_foreground_wmap} large differences
in the component which is considered as foreground
from the 1yr to the 3yr data.
The signal which is considered as foreground is much smaller in the 3yr data
than in the 1yr data.
This change can be compared with the average difference
$\Delta_{\hbox{\scriptsize 3yr}}^{\hbox{\scriptsize 1yr}} :=
\langle \delta T(i)_{\hbox{\scriptsize pri}}^{\hbox{\scriptsize 1yr}}-
\delta T(i)_{\hbox{\scriptsize pri}}^{\hbox{\scriptsize 3yr}}
\rangle^{\hbox{\scriptsize Kp2}}$
between the 1yr and 3yr data which is also given in
Table \ref{Tab:DA_foreground_wmap}.
An order of magnitude estimate of the foreground uncertainty may thus
lie in the range $20\dots 30 \mu\hbox{K}$.

Because of these uncertainties
we conclude that the analysis in \cite{Cornish_Spergel_Starkman_Komatsu_2003}
does not rule out a  toroidal universe.

\section{Probabilities from the covariance matrix}

\begin{figure}[htb]
\begin{center}
\begin{minipage}{9cm}
\vspace*{-20pt}\includegraphics[width=9.0cm]{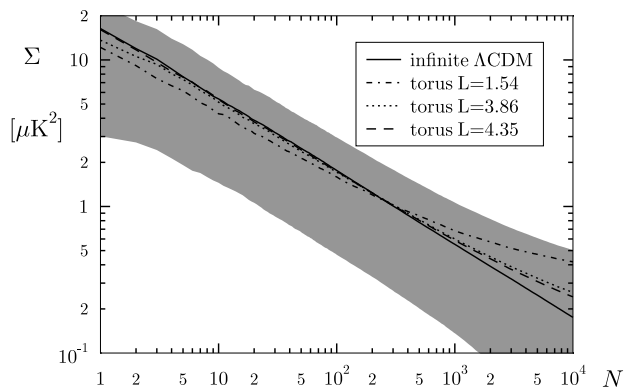}
\end{minipage}
\vspace*{-20pt}
\end{center}
\caption{\label{Fig:Covariance_non_diag}
The average $\Sigma$ of the non-diagonal elements is shown in
dependence on the ensemble size $N$ for which the average is computed.
The grey band displays the average deviation for the $L=3.86$ torus ensemble.
The $1/\sqrt N$ decay in the case of the infinite $\Lambda$CDM model is
in this logarithmic plot revealed by the straight line with slope $-1/2$.
The finite models deviate the stronger the smaller the torus length is
in order to approach non-vanishing diagonal elements.
}
\end{figure}

In \cite{Phillips_Kogut_2006} the covariance matrix
$C_{lm}^{l'm'} := \left< a_{lm} a_{l'm'}^\star\right>$
for the torus universe is compared with the WMAP data
and a 95\% confidence limit is found that the torus length is $L>4$.
(Here $\left< \dots \right>$ denotes an ensemble average.)
The covariance matrix for the infinite model
$C_{lm}^{l'm'} = C_l \delta_{ll'} \delta_{mm'}$ being diagonal
contrasts to the corresponding one of an anisotropic multiply connected space
having non-vanishing non-diagonal elements.
These can thus serve as a fingerprint for an anisotropic topology
of the universe.
However, this statement refers to the ensemble average.
For an individual model non-diagonal elements are observed in both cases.
In Fig.\ \ref{Fig:Covariance_non_diag} the average
$\Sigma :=
\left<\, \left| \Big<C_{lm}^{l'm'} \Big>_N \right|\, \right>_{(lm)\neq(l'm')}$
over $N$ models of the non-diagonal elements is shown
for several torus models and the infinite $\Lambda$CDM model.
The grey band shows the average deviation from the mean  value for
the simulations belonging to the $L=3.86$ torus.
Here and in the following, the covariance matrix is truncated at
$l_{\hbox{\scriptsize max}}=30$.
Numerical tests based on sky maps of the torus model show
that for side lengths around $L\simeq 4$ the modes above $l\simeq 20$
do not improve the likelihood to detect a torus topology.
Therefore, the restriction to $l\leq 30$ in the covariance matrix is justified.
For the infinite model, the average decays as $1/\sqrt N$ leading
asymptotically to vanishing non-diagonal elements.
Fig.\ \ref{Fig:Covariance_non_diag} shows
that the torus models display indeed a deviation from the $1/\sqrt N$ decay
leading to non-vanishing non-diagonal elements in the covariance matrix.
To reveal this behaviour, an average over a large number of models
is necessary, but we have observations only from one Universe.
In addition, this is aggravated by the large average deviation
as shown in Fig.\ \ref{Fig:Covariance_non_diag}.
A closer inspection reveals that the non-vanishing non-diagonal elements
in the case of the torus models are due to their real parts,
whereas the imaginary parts vanish according to the $1/\sqrt N$ decay.

\begin{figure}[ttt]
\begin{center}
\begin{minipage}{9cm}
\vspace*{-30pt}\includegraphics[width=9.0cm]{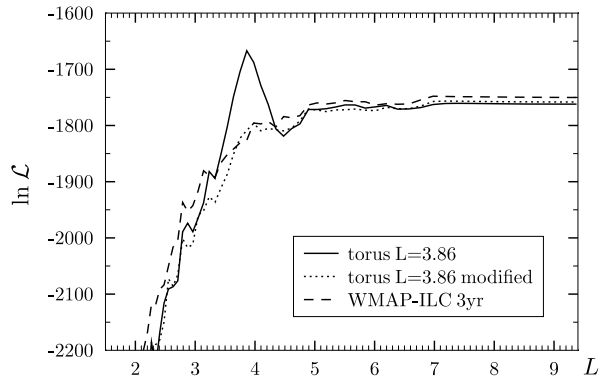}
\end{minipage}
\vspace*{-30pt}
\end{center}
\caption{\label{Fig:Likelihood}
The logarithm of the likelihood $\ln{\cal L}$ is shown for the
covariance matrix $C_{ss'}$ belonging to the torus model
with length $L$.
The $\chi^2$ is computed using the $a_s$ belonging to
the torus model with $L=3.86$ (full curve) and for the $a_s$
obtained from the WMAP-ILC 3yr map (dashed curve).
The sharp maximum in the full curve clearly ``detects'' the
underlying torus model.
However, modifying the phases destroys the maximum (dotted curve).
}
\end{figure}

The covariance matrix can nevertheless be used to detect
an anisotropic topology by computing the logarithm of the likelihood
$\ln {\cal L} = -\frac 12 ( \ln \det C_{ss'} + \chi^2) + \hbox{const.}$,
where $s := l(l+1)+m$ and
$\chi^2 = \min \sum_{s,s'} a_s^\star C_{ss'}^{-1} a_{s'}$.
The minimum is obtained by evaluating the sum $\sum_{s,s'}$ for all  rotations of $a_s$,
i.\,e.\ for all possible orientations of the fundamental cell with respect to
the CMB sky.
The method of detecting a toroidal universe in this way
is discussed in \cite{Kunz_et_al_2006, Kunz_et_al_2007}.
In Fig.\ \ref{Fig:Likelihood} we show $\ln {\cal L}$ for the
covariance matrix of the torus model in dependence on the length $L$
computed from three different sets of $a_{lm}\equiv a_s$.
One set $\{a_s\}$ is computed using a torus model with $L=3.86$ (full curve)
and provides the largest likelihood for the covariance matrix
at $L=3.86$ thus revealing the topology.
No such maximum is observed using the $a_s$ obtained from the
WMAP-ILC 3yr sky map (without mask) as shown by the dashed curve.
Does this exclude a toroidal topology around $L\simeq 4$?
The answer depends on the accuracy of the $a_s$ obtained from
the WMAP data.
This accuracy in turn is determined by the accuracy of the ILC sky map,
which we have discussed in the last section.
To illustrate this point, we modify the phases of the $a_s$ of
the torus model with $L=3.86$ by
$a_s \rightarrow a_s \cdot e^{\hbox{\scriptsize i} r_s}$,
where $r_s$ is a Gaussian random variable with unit variance.
This blurs the phases by $\simeq 57^\circ$.
The result is shown as the dotted curve in Fig.\ \ref{Fig:Likelihood}
and it is seen that this inaccuracy in the phases destroys the
peak at $L=3.86$.
Such an inaccuracy in the phases would make the detection of
an anisotropic topology impossible.
The accuracy of the phases is discussed in
\cite{Naselsky_Verkhodanov_Nielsen_2007,Chiang_Naselsky_Coles_2007}
and the WMAP team itself warns of using the ILC map for such a
cosmological analysis.
In \cite{Naselsky_Verkhodanov_Nielsen_2007} it is shown
that the Minimal Variance Method used in the construction of the ILC map
leads to a wrong sign for $\sim 40\%$ of the coefficients $a_{20}$
from $10^4$ Monte Carlo simulations, i.\,e.\ a phase shift of $180^\circ$.
The assumption that the CMB is a Gaussian random field leads to phases
which are uniformly distributed over the interval $[0^\circ,360^\circ]$.
In \cite{Chiang_Naselsky_Coles_2007} it is shown
that the phases for $l\leq 10$ are consistent with a
uniform distribution but not the differences between the phases
which should also be uniformly distributed.
This should be taken as a warning that the phases might not be sufficiently accurately
determined by the ILC map.

\begin{figure}[htb]
\begin{center}
\begin{minipage}{18cm}
\begin{minipage}{9cm}
\includegraphics[width=9.0cm]{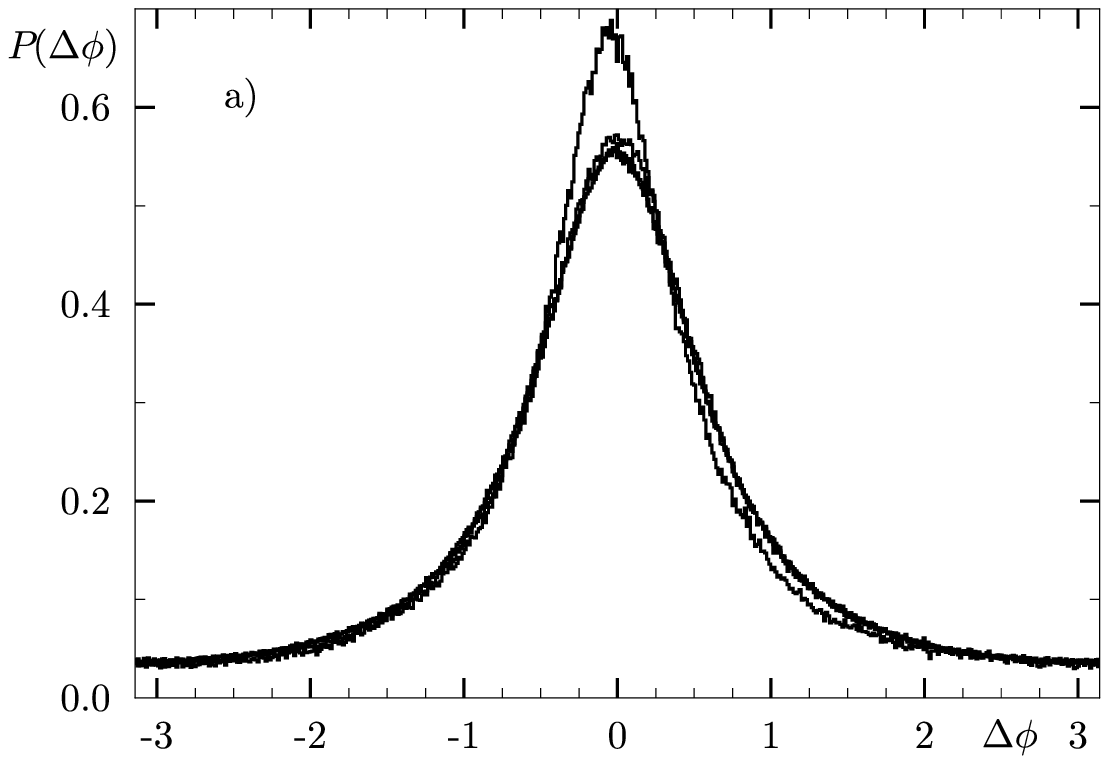}
\end{minipage}
\begin{minipage}{9cm}
\hspace*{-30pt}\includegraphics[width=9.0cm]{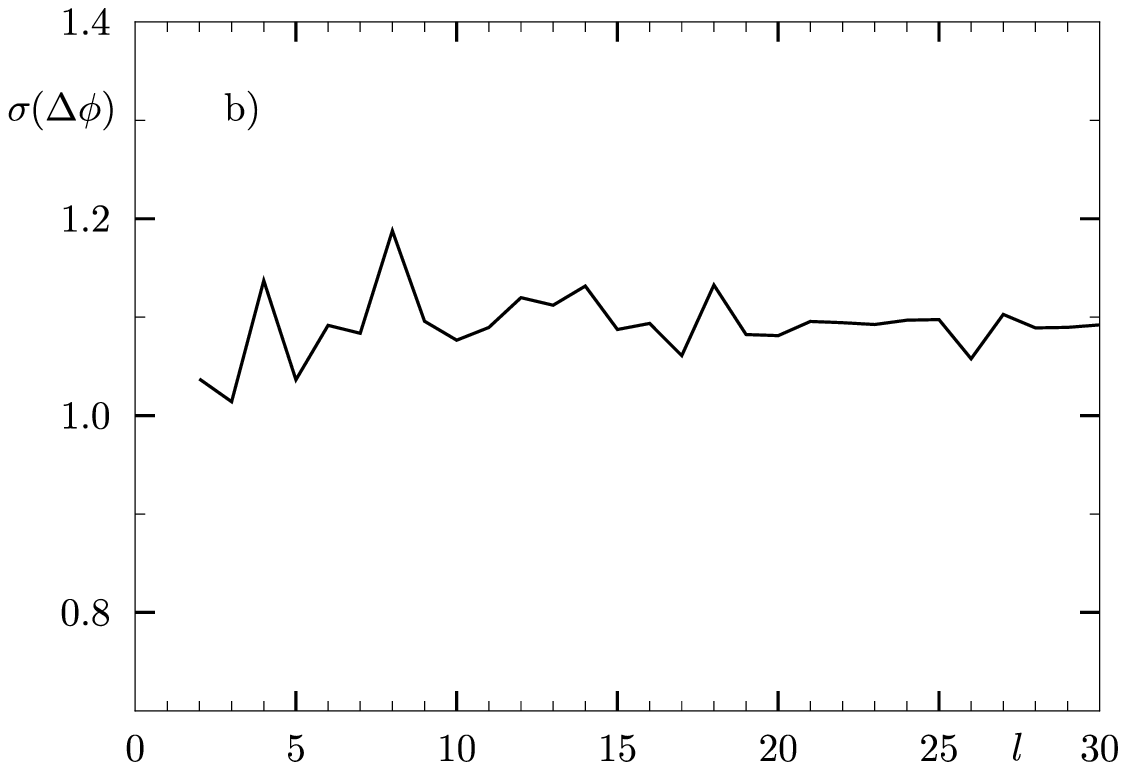}
\end{minipage}
\end{minipage}
\end{center}
\caption{\label{Fig:phase_error}
Panel a) shows the distribution of the error $\Delta\phi$ of the phases
(in radian)
by replacing the torus temperature fluctuations within the kp0 mask
by those of an infinite $\Lambda$CDM model for $l=5$, 10, 20, and 30.
The distribution with the highest peak at $\Delta\phi=0$ belongs to $l=5$
whereas the other distributions are indistinguishable.
Panel b) shows that the standard deviation $\sigma(\Delta\phi)$
is $\sigma(\Delta\phi) \simeq 1.1$ independent of $l$.
}
\end{figure}

Since the accuracy of the phases is unknown,
we try to obtain an order of magnitude estimate by the following
procedure.
Assuming that the ILC map displays outside the kp0 mask the genuine CMB signal,
i.\,e.\ without any foregrounds, which is surely not true,
and within the kp0 mask a strongly disturbed signal
which eradicates the CMB signal,
we construct sky maps which have outside the kp0 mask the corresponding part
of a torus simulation and within the mask a map obtained from
an infinite $\Lambda$CDM simulation.
In this way we combine 950 torus sky maps with 100 infinite $\Lambda$CDM sky maps
leading to 95\,000 data sets of the spherical expansion coefficients $a_{lm}$.
Comparing the phases of these sky maps with the phases of the
corresponding torus simulations without mask
reveals that the phase change $\Delta\phi$ has a standard deviation
$\sigma(\Delta\phi) \simeq 63^\circ$.
The large deviations which are of the same order as used in Fig.\,\ref{Fig:Likelihood},
have their roots in the strong oscillatory nature of
the spherical harmonics which leads to many cancellations in the integration
required for the computation of the $a_{lm}$'s.
In Fig.\,\ref{Fig:phase_error}a the distributions of the phase changes $\Delta\phi$
are shown for four different values of $l$, i.\,e.\ $l=5$, 10, 20, and 30.
The distributions are very similar and have a standard deviation
$\sigma(\Delta\phi)$ independent of $l$ as revealed by Fig.\,\ref{Fig:phase_error}b.
This demonstrates that the accuracy of the phases might be low enough to
hide a signature of the topology.

Since the true accuracy of the phases is unknown,
one cannot draw a final conclusion whether the covariance matrix signature
already rejects a toroidal topology for the Universe.
Notice that the correlation function $C(\vartheta)$ and the
angular power spectrum $\delta T_l^2$ do not depend on these phases
such that the very low power at angles $\vartheta \gtrsim 60^\circ$
is a much more robust observation pointing to a non-trivial topology.

The signature of the non-diagonal elements of the covariance matrix
$C_{lm}^{l'm'}$ is not only revealed in the spherical harmonics basis
but also directly in the pixel space $C(\hat n,\hat n')$.
This alternative is introduced in
\cite{Bond_Pogosyan_Souradeep_1999b}
and applied to compact hyperbolic spaces. 
In \cite{Inoue_Sugiyama_2003} a Bayesian analysis is carried out for
the torus model using the pixel space correlation $C(\hat n,\hat n')$
based on the COBE-DMR data.
It is found that for $\varepsilon =L/(2\Delta\eta)\geq 0.6$, i.\,e.\ $L\geq 4$,
the torus models are consistent with the standard $\Lambda$CDM model
at the $2\sigma$ level, and for some limited choices of orientations
the fit to the COBE data is considerably better than that of the
infinite model.
Such a Bayesian analysis based on $C(\hat n,\hat n')$ using the WMAP 3yr
data is not carried out in this paper
such that the advantage of the torus model over the $\Lambda$CDM model
with respect to this measure remains open here.

\section{Summary}

Our analysis shows that the simplest non-trivial flat topology,
the cubic torus with a side length smaller than the diameter of
the decoupling surface, is compatible with the WMAP 3-year data,
and describes these data much better than the standard $\Lambda$CDM
model.
It is very intriguing that the toroidal topology provides also the simplest
way to solve the problem of initial conditions for the low-scale
inflation \cite{Linde_2004}.

CMBFAST (www.cmbfast.org) and HEALPix (healpix.jpl.nasa.gov)
\cite{Gorski_Hivon_Banday_Wandelt_Hansen_Reinecke_Bartelmann_2005}
were used in this work.

\section*{References}

\bibliography{../../bib_astro}

\bibliographystyle{h-physrev3}

\end{document}